\pdfoutput=1
\documentclass[aps,prd,twocolumn,superscriptaddress,nofootinbib,floatfix]{revtex4-2}
\usepackage{amsmath,amssymb}
\usepackage{graphicx}
\usepackage{booktabs}
\usepackage[hidelinks]{hyperref}

\newcommand{\Msun}{M_\odot}
\newcommand{\Gyr}{\,\mathrm{Gyr}}
\newcommand{\Myr}{\,\mathrm{Myr}}
\newcommand{\Mpc}{\,\mathrm{Mpc}}
\newcommand{\eV}{\,\mathrm{eV}}

\newcommand{\aap}{Astron. Astrophys.}
\newcommand{\mnras}{Mon. Not. Roy. Astron. Soc.}

\begin{document}

\title{\texorpdfstring{\Large A $\nu$ mechanism for speeding up star formation: \\ 
from the first galaxies to the Local Group}{A nu mechanism for speeding up star formation: from the first galaxies to the Local Group}}

\author{Wen~Yin}
\affiliation{Department of Physics, Tokyo Metropolitan University, Minami-Osawa, Hachioji-shi, Tokyo 192-0397, Japan}

\author{Daisuke~Taniguchi}
\thanks{The Tokyo Center For Excellence Project, Tokyo Metropolitan University.}
\affiliation{Department of Physics, Tokyo Metropolitan University, Minami-Osawa, Hachioji-shi, Tokyo 192-0397, Japan}
\affiliation{Universit\'e C\^ote d'Azur, Observatoire de la C\^ote d'Azur, CNRS, Laboratoire Lagrange, Bd de l'Observatoire, CS 34229, 06304 Nice cedex 4, France}

\author{Tatsuki~Sato}
\affiliation{Department of Physics, Tokyo Metropolitan University, Minami-Osawa, Hachioji-shi, Tokyo 192-0397, Japan}

\author{Yuma~Narita} \affiliation{Department of Physics, Tokyo Metropolitan University, Minami-Osawa, Hachioji-shi, Tokyo 192-0397, Japan} 

\author{Kyosuke~S.~Sato}
\affiliation{Astronomical Science Program, The Graduate University for Advanced Studies, SOKENDAI, 2-21-1 Osawa, Mitaka, Tokyo 181-8588, Japan}
\affiliation{National Astronomical Observatory of Japan, 2-21-1 Osawa, Mitaka, Tokyo 181-8588, Japan}

\author{Itsuki~Ogami}
\affiliation{Department of Physics, Tokyo Metropolitan University, Minami-Osawa, Hachioji-shi, Tokyo 192-0397, Japan}
\affiliation{Department of Advanced Sciences, Faculty of Science and Engineering, Hosei University, 3-7-2 Kajino-cho, Koganei, Tokyo 184-8584, Japan}

\author{Junichi~Baba}
\affiliation{Amanogawa Galaxy Astronomy Research Center, Graduate School of Science and Engineering, Kagoshima University, 1-21-35
Korimoto, Kagoshima 890-0065, Japan}
\affiliation{National Astronomical Observatory of Japan, 2-21-1 Osawa, Mitaka, Tokyo 181-8588, Japan}

\date{\today}

\begin{abstract}
The James Webb Space Telescope finds more UV-bright galaxies at $z\gtrsim12$ than expected, and enhanced star formation about $2\Gyr$ ago has been reported in several galaxies of the Local Group.
Star formation is regulated by feedback, in which core-collapse supernovae, powered by neutrinos, play a central role, and simulations of the progenitor of SN~1987A find that mixing between the electron neutrino and an eV-scale sterile neutrino makes it fail to explode.
We point out a new mechanism in which a light scalar field controlling this mixing can speed up star formation at particular times or places, without injecting energy.
As a demonstration, we show that in a minimal model with degenerate vacua, the excess reported by JADES is accounted for by the early stage of the domain walls, when the mixing is on everywhere, and in the late Universe, when a wall crosses the Local Group, the star-formation rate of all its gas-rich galaxies rises at once, which may explain these enhancements.
The general prediction is that the walls leave thin surfaces with fewer core-collapse supernovae and faster-growing galaxies on our past light cone, which is consistent with present supernova samples and can be probed by the Legacy Survey of Space and Time.
If the scalar also couples to photons, it can explain the hint of isotropic cosmic birefringence at more than $4\sigma$.
\end{abstract}

\maketitle
\section{Introduction}
Two sets of observations suggest episodes in which stars formed faster than expected.
First, the James Webb Space Telescope (JWST) finds more UV-bright galaxies at redshift $z\gtrsim12$ than predicted by pre-JWST models~\cite{Whitler2025,Donnan2024,Weibel2025}, an excess that grows with redshift (Fig.~\ref{fig:uvlf}); proposed explanations include feedback-free starbursts~\cite{Dekel:2023ddd}, weak dust attenuation~\cite{Ferrara:2022dqw} and bursty star formation~\cite{Mason:2022tiy,Shen:2023cva}.
Second, the star-formation histories of the Local Group (the Milky Way, M31 and their neighbours), reconstructed from colour--magnitude diagrams of resolved stars, show an enhanced star-formation rate (SFR) about $2\Gyr$ ago in the Milky Way~\cite{RuizLara2020,Mor2019} and peaks at similar look-back times in several other galaxies (see Fig.~\ref{fig:sfh}); these are usually attributed to processes specific to each galaxy or pair, and part of the coincidence may be systematic (Sec.~\ref{sec:ism}).
Both may have astrophysical explanations, but they could also hint at physics beyond the Standard Model (BSM) that acts on star formation, which will become important if further observations make the phenomena more robust. 

How fast gas turns into stars is known to be regulated by feedback from massive stars, in which core-collapse (CC) supernovae (SNe), the explosions that end the lives of stars heavier than about $8\,\Msun$, play a central role.
In the first galaxies this feedback limits the fraction of the accreted gas that turns into stars~\cite{Dekel:1986ehj}.
In a galactic disc the weight of the gas is supported by the pressure in the midplane, whose thermal, turbulent and magnetic parts are supplied by feedback in proportion to the SFR surface density~\cite{Ostriker:2010zc,Ostriker:2011jb,Kim:2011hj,Kim:2013xaa,OstrikerKim2022}; the turbulent part is sustained by the momentum injected by SNe, mostly CC ones~\cite{KimOstriker2015,Bacchini2020,Meng2026}.
If CC SNe stopped exploding for a while, the gas would lose part of this support and form stars faster, without any energy being injected.

It has been found that new neutrino physics can stop CC SNe.
A CC SN is powered by neutrinos: when the iron core of the star collapses into a proto-neutron star, the outgoing shock wave stalls, and in the standard picture it is revived by electron neutrinos $\nu_e$ and antineutrinos $\bar\nu_e$ from the proto-neutron star, which heat the matter behind it~\cite{Burrows:1993pi,Janka:2012wk}.
A sterile neutrino, a neutral fermion that does not feel the weak interaction, can take this heating away.
If $\nu_e$ mixes with a sterile neutrino of mass $\simeq1$--$2\eV$, the matter of the SN partly converts $\nu_e$ and $\bar\nu_e$ resonantly into sterile states near the region where they would heat the gas (the Mikheyev--Smirnov--Wolfenstein, or MSW, effect)~\cite{Nunokawa:1997ct,Wu:2013gxa}, and simulations of the progenitor of SN~1987A find that the shock is not revived for large enough mixing~\cite{Mori:2025lky}.
Such mixing cannot be efficient in the galaxies where CC SNe are observed today, nor in the Large Magellanic Cloud when SN~1987A exploded; the observed explosions thus limit the mixing~\cite{Mori:2025lky}. In addition, the mixing is constrained by early-universe cosmology (see, e.g., Ref.~\cite{Hannestad:2012ky}). 

In this paper we point out three things:
\begin{itemize}
\item BSM neutrino physics can control star formation: mixing of $\nu_e$ with an eV-scale sterile neutrino removes the feedback of CC SNe, so that gas forms stars faster without any energy being injected.
\item Star-formation observables, such as the UV luminosity functions of the first galaxies, star-formation histories and the rates of CC SNe, thus become probes of the neutrino sector.
\item The limits from the observed SNe and from early-universe cosmology are evaded if the mixing is controlled by a light scalar field.
\end{itemize}

To illustrate them, we build a minimal model and apply it, as a trial, to the first galaxies and to the Local Group.
The potential of the scalar has two degenerate minima (vacua), in which the mixing vanishes. 
From around recombination, however, the finite-density effect of the relic neutrinos drives the scalar from the top of the potential, where it sat with negligible mixing before, to  temporary minima with non-vanishing mixings, so that the CC SNe of the first galaxies fail; they resume once the relic neutrinos are diluted by the expansion of the Universe, which accounts for the excess found by the JWST Advanced Deep Extragalactic Survey (JADES)~\cite{Whitler2025} (Sec.~\ref{sec:uvlf}).
Domain walls~\cite{Zeldovich:1974uw} form between regions displaced towards different vacua, and today the mixing is on only inside them: a wall is Mpc thick, stops CC SNe in a galaxy for $\sim 0.1\Gyr$ and sweeps the Local Group within $\sim10$--$20\Myr$.
The walls leave on our past light cone thin surfaces of galaxies in which CC SNe are reduced while SNe~Ia, the thermonuclear explosions of white dwarfs, are normal; present supernova samples are consistent with such surfaces, and the Legacy Survey of Space and Time (LSST)~\cite{LSST:2008ijt} can reveal them.
If the scalar also couples to photons, as an axion-like particle does, the same evolution rotates the polarization of the cosmic microwave background (CMB) and can account for the $>4\sigma$ hint of isotropic cosmic birefringence~\cite{Minami:2020odp,Diego-Palazuelos:2022dsq,Eskilt:2022wav,Cosmoglobe:2023pgf,Eskilt:2026imm}, for which walls of axion-like fields are known as a natural potential source~\cite{Takahashi:2020tqv,Kitajima:2022jzz,Gonzalez:2022mcx,Ferreira:2023jbu,Lee:2025yvn}.
Walls that instead change parameters of the Standard Model have a set of severe constraints that our model does not have~\cite{To_appear_Sato_et_al}.

\section{Model}
\begin{figure*}[t]
\includegraphics[width=\textwidth]{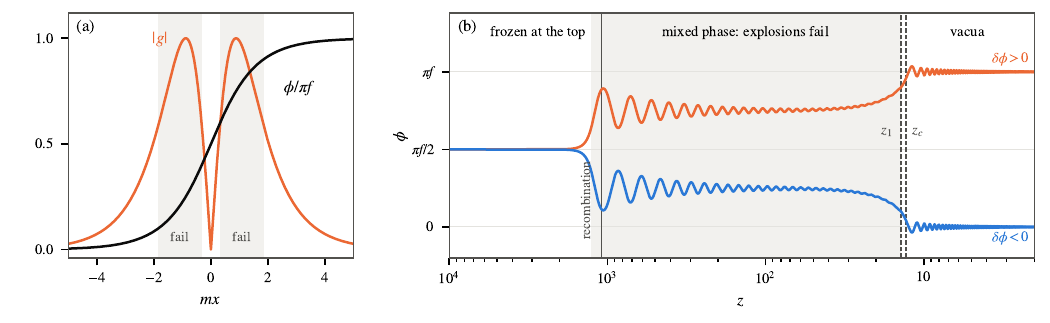}
\caption{\label{fig:field} (a) Field $\phi/\pi f$ (black) and mixing strength $|g|$ (orange) across the benchmark wall, as functions of the distance $x$ from its centre, the top of $V$. Explosions fail in the two shaded intervals, where $|X|\geq0.45\eV^2$. (b) History of the field in two patches with opposite signs of $\delta\phi$ (benchmark, normal ordering, $m_1=1.0$~meV): frozen at the top of $V$ until shortly before recombination (vertical line), then in the mixed phase on either side, where explosions fail (shaded) until $z_1$, and in the vacua about $20\Myr$ after $z_c$, at which the minimum induced by the relic neutrinos reaches them (dashed lines); walls form between such patches. }
\end{figure*}

Consider a real periodic scalar $\phi\equiv\phi+2\pi f$ and two gauge-singlet fermions $N_R$ and $S_L$ that carry lepton number with the Lagrangian\footnote{If $S_L$ had a Majorana mass above about $1$~meV, or if $N_R$ were a Majorana fermion without $S_L$, the active neutrinos would instead become heavier where the mixing is on; in the latter case the relic neutrinos would hold the field at the top of $V$ until $z\simeq0.6$--$0.7$ and drag the walls.}
\begin{align}
\mathcal L ={}& \tfrac12(\partial\phi)^2-V(\phi) \nonumber\\
&+\bar N_R i\partial\!\!\!/ N_R+\bar S_L i\partial\!\!\!/ S_L \nonumber\\
&-\Big[M\,\bar S_L N_R+y_0\,g(\phi)\,\bar L\tilde H N_R+\mathrm{h.c.}\Big],
\label{eq:L}
\end{align}
with\footnote{The suppression of the Coleman--Weinberg correction can be managed with additional symmetries such as that of Ref.~\cite{Hook:2018jle}. See also Ref.~\cite{Yin:2026ova} for how the neutrino mass is explained in a similar model.}
\begin{equation}
g(\phi)=\sin\frac{2\phi}{f},\qquad
V(\phi)=V_{\rm top}\sin^2\frac{\phi}{f},
\label{eq:Vg}
\end{equation}
where $V_{\rm top}\equiv m^2f^2/2$, so that $m$ is the mass in the vacua.
Here $L$ is the lepton doublet that contains $\nu_{eL}$, $H$ is the Higgs doublet, with $\tilde H=i\sigma_2H^*$ and vacuum expectation value $v/\sqrt2$ ($v=246\,\mathrm{GeV}$), and $N_R$ and $S_L$ are sterile neutrinos.
The functions $g$ and $V$ are the lowest harmonics allowed by (electron) lepton number, which the active neutrino mass term below breaks, and two parities, 
\begin{align} P&: \phi\to-\phi,\ N_R\to -N_R,\ S_L\to -S_L,\\  R&: \phi\to\pi f-\phi,\ N_R\to -N_R,\ S_L\to -S_L.
\end{align}
The field $\phi$ has two vacua, $\phi=0$ and $\pi f$, in which $g=0$; the two sterile fermions form one particle of mass $M$, which mixes with $\nu_e$ through the Higgs only where $g\neq0$, i.e.\ inside the walls between the vacua (except at their centre).
The vacua $\phi=k\pi f$ are separated by walls $\phi=2f\arctan e^{mx}$, where $x$ is the distance from the centre of the wall, of thickness $\sim m^{-1}$ and tension $\sigma=2mf^2$.
$P$ is unbroken in the vacua and $R$ at the top of $V$, $\phi=\pi f/2$, which is the centre of the wall, so the mixing vanishes at both by symmetry (no mixing in the vacua); across the wall $|g|=2|\tanh mx|/\cosh mx$, which reaches $1$ on either side of the centre [Fig.~\ref{fig:field}(a)].
With $m_D=g\,y_0 v/\sqrt2$, $N_R$ pairs with $\cos\theta\,S_L+\sin\theta\,\nu_{eL}$, $\tan\theta=m_D/M$, into a Dirac fermion; the $\nu_e$--sterile system has $\sin2\theta=2m_DM/(M^2+m_D^2)$ and $\delta m^2=M^2+m_D^2$, so that
\begin{equation}
X\equiv\delta m^2\sin2\theta = 2Mm_D
\label{eq:X}
\end{equation}
is exactly linear in $g$.
At $|g|=1$ the benchmark (Table~\ref{tab:constraints}) has $\delta m^2=3.9\eV^2$ and $\sin^22\theta=0.04$, i.e.\ $X=0.78\eV^2$, the parameters of one of the simulated models without shock revival~\cite{Mori:2025lky}.
In these simulations explosions fail for $|X|\geq0.45\eV^2$ and are delayed and weaker for smaller $|X|$; we describe this by the fraction $S$ of standard CC-SN feedback, from $S=0$, i.e.\ CC SNe fail, for $|X|\geq0.45\eV^2$ to $S=1$ for $|X|\leq0.05\eV^2$ (Sec.~\ref{sec:map}).

The active neutrinos get Majorana masses $m_i$ from the operator $(\overline{L^c}\tilde H^*)(\tilde H^\dagger L)/\Lambda$, while the sterile pair has no Majorana mass, so that the combination of $\nu_{eL}$ and $S_L$ that does not pair with $N_R$ gets no mass from it; where the mixing is on, the active neutrinos are thus partly $S_L$ and become slightly lighter, $m_i\to m_i(1-\sin^2\theta\,|U_{ei}|^2)$, by $\sim10^{-5}\eV$.

\begin{figure}[t]
\includegraphics[width=\columnwidth]{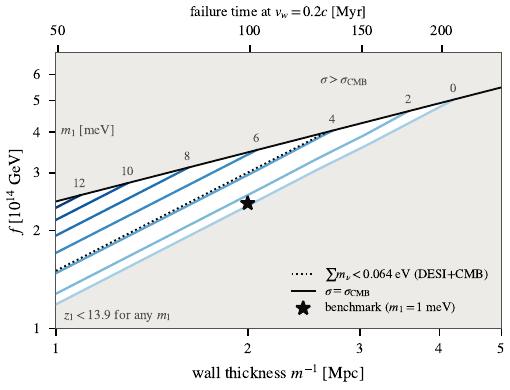}
\caption{\label{fig:region}Allowed region (white) of the decay constant $f$ and the thickness $m^{-1}$ of the walls. Each blue line, $f=\sqrt{2V_{\rm top}}/m$, has the $V_{\rm top}$ for which the CC SNe resume at $z_1=13.9$ for the lightest neutrino mass $m_1$ indicated (in meV); this $V_{\rm top}$ does not depend on $m$. Below the line $m_1=0$ the CC SNe resume later for any $m_1$, and above the black line $\sigma>\sigma_{\rm CMB}$~\cite{Sousa:2015cqa}. Dotted: $\sum m_\nu<0.064$~eV from DESI and the CMB~\cite{DESI:2025zgx}; the CMB alone, $\sum m_\nu<0.24$~eV~\cite{Planck:2018vyg}, allows the whole region shown. Star: the benchmark (Table~\ref{tab:constraints}). Top axis: the time for which a passing wall stops the CC SNe of a galaxy at $v_w=0.2c$.}
\end{figure}

The parameters are related as follows.
The mass $m$ sets the thickness of the walls, $m^{-1}$, their tension, $\sigma=4V_{\rm top}/m$, and $f=\sqrt{2V_{\rm top}}/m$; the bound on domain-wall networks from the CMB, $\sigma<\sigma_{\rm CMB}$~\cite{Sousa:2015cqa}, requires $m^{-1}\lesssim4\Mpc\times(3\times10^{-13}\eV^4/V_{\rm top})$.
The thickness also sets how long a passing wall stops the CC SNe of a galaxy, $\propto m^{-1}/v_w$ for a wall speed $v_w$. The epoch $z_1\simeq14$ at which CC SNe resume fixes $V_{\rm top}$ for a given mass $m_1$ of the lightest neutrino, independently of $m$, and Fig.~\ref{fig:region} shows the allowed region; in it the fit to JADES depends only on $z_1$, and the average rotation of the CMB polarization does not depend on $f$, $m$ and $\sigma$ in the range of interest (Sec.~\ref{sec:biref}, cf.~Ref.~\cite{Takahashi:2020tqv}). Thus the parameter region explaining JADES and cosmic birefringence is large. 

Within this region we take $m^{-1}=2\Mpc$, for which a passing wall stops the CC SNe for $\sim0.1\Gyr$ at $v_w=0.2c$, and $\sigma\simeq0.5\,\sigma_{\rm CMB}$, which is chosen to satisfy the constraint from the domain-wall problem, $\sigma<\sigma_{\rm CMB}$, with a margin.
Table~\ref{tab:constraints} lists the benchmark, in which $m_1=1.0$~meV gives $z_1\simeq13.9$, close to the best fit to JADES (Sec.~\ref{sec:uvlf}), and the constraints. Any initial displacement $\delta\phi_{\rm rms}$ from $10^{-8}f$ to $5\times10^{-4}f$, the largest allowed by the CMB (Sec.~\ref{sec:early}), changes $z_1$ and the later epochs by less than $0.4$; the benchmark has $10^{-7}f$.

\begin{table}[t]
\caption{\label{tab:constraints}Benchmark and constraints. 
``inside'' denotes when the wall passes by.}
\begin{ruledtabular}
\begin{tabular}{p{0.36\columnwidth}p{0.58\columnwidth}}
\multicolumn{2}{l}{\textit{Benchmark}} \\ \hline
Sterile pair & $M=1.96\eV$, $y_0=1.14\times10^{-12}$ \\
Scalar & $f=2.4\times10^{14}$~GeV, $m^{-1}=2.0\Mpc$ ($m=3.2\times10^{-30}\eV$), $V_{\rm top}=3.0\times10^{-13}\eV^4$ \\
Wall & $v_w=0.2c$, $\sigma=3.8\times10^{17}\eV^3=0.5\,\sigma_{\rm CMB}$ \\
Neutrinos & Majorana, normal ordering, $m_1=1.0$~meV \\
Initial field & $\delta\phi_{\rm rms}=10^{-7}f$ from the top of $V$ \\ \hline
\multicolumn{2}{l}{\textit{Constraints}} \\ \hline
Laboratory, reactor, KATRIN, clocks & residual $\sin^22\theta\lesssim5\times10^{-7}$ today; no $\alpha$/$m_e$ shift \\
SN~1987A & LMC was in the post-wall vacuum in 1987 \\
$^{187}$Re, Oklo & only $\sin^2\theta\,(M/Q)^2\sim10^{-8}$ shift of $\beta$ rates inside ($Q$: decay energy) \\
Sun (inside) & $\lesssim2\%$ of $\nu_e\to$ sterile; no structural effect \\
BBN & $\Delta N_{\rm eff}\sim10^{-11}$ \\
Relic neutrinos & $\epsilon\simeq2\times10^{-4}$ at $z=0.15$: no drag or distortion of the wall \\
\end{tabular}
\end{ruledtabular}
\end{table}

\section{Cosmological history}
\subsection{Relic neutrinos and the early Universe}
While $H\gg m$ the field is frozen at a displacement $\delta\phi$ from the top of $V$, $\phi=\pi f/2+\delta\phi$, where $g\simeq-2\delta\phi/f$.
We take $\delta\phi$ to vary in space with zero mean and a small rms $\delta\phi_{\rm rms}$ (Table~\ref{tab:constraints}). The top is a symmetry-enhanced point, where $R$ is unbroken; the field can be placed there during inflation through mixing with the inflaton or by thermal effects~\cite{Daido:2017wwb,Takahashi:2019pqf, Gonzalez:2022mcx}.
The mixing in the early Universe is then suppressed ($\sin^22\theta\sim10^{-15}$), and non-resonant production~\cite{Dodelson:1993je,Hannestad:2012ky} leaves a negligible abundance of sterile states (Table~\ref{tab:constraints}; Sec.~\ref{sec:early}, see also Ref.~\cite{Farzan:2019yvo}).
The later history is set by the relic neutrinos via the mixing.
Since $\phi$ changes over a Hubble time or across a Mpc-thick wall, at rates $\lesssim10^{-28}\eV$ far below the splitting $\simeq M$ between the active and sterile states, they follow the active states adiabatically, without conversion into sterile states, and the active-neutrino masses drop to $m_i(1-\sin^2\theta\,|U_{ei}|^2)$ where the mixing is on, which lowers the energy density of the relic neutrinos by
\begin{equation}
\Delta E_\nu=\sin^2\theta\sum_i|U_{ei}|^2m_i^2\int\!\frac{d^3p}{(2\pi)^3}\,\frac{2f_\nu(p)}{\sqrt{p^2+m_i^2}},
\label{eq:Enu}
\end{equation}
where $f_\nu$ is the relic distribution; for nonrelativistic neutrinos it is the number density times the mass shift, $\Delta E_\nu\simeq\sin^2\theta\,n_\nu\sum_i|U_{ei}|^2m_i$ with $n_\nu=112\,(1+z)^3\,\mathrm{cm^{-3}}$.
This term is proportional to $g^2$ to a good approximation, so the field feels
\begin{equation}
V_{\rm eff}\simeq V_{\rm top}\Big[\sin^2\frac{\phi}{f}-\epsilon(z)\sin^2\frac{2\phi}{f}\Big], ~~\epsilon\equiv \Delta E_\nu/V_{\rm top}
\label{eq:Veff}
\end{equation}
where $\epsilon$ is evaluated at $|g|=1$.
The relic neutrinos make the top more unstable and, for $\epsilon>1/4$, also the vacua; the minimum is then the mixed phase $\cos(2\phi/f)=1/(4\epsilon)$, in which $|X|\geq0.45\eV^2$ for $\epsilon>0.31$.
For the benchmark with normal mass ordering and $m_1=1.0$~meV, with which the explosions resume when the luminosity functions require (below), a numerical FRW integration, in which patches with different $\delta\phi$ are evolved as separate homogeneous regions (Sec.~\ref{sec:early}), gives the following history.
The field leaves the top at $z\simeq1200$--$1530$, shortly before recombination, and settles in the minimum induced by the relic neutrinos (the mixed phase) on the side given by the sign of the local $\delta\phi$, so that walls already form then, between patches on opposite sides [Fig.~\ref{fig:field}(b)].

These walls make up a network which, after a transient, is close to scaling~\cite{Press:1989yh,Garagounis:2002kt,Hiramatsu:2013qaa,Martins:2016ois,Kitajima:2022jzz,Gonzalez:2022mcx,Kitajima:2023kzu,Kitajima:2026llw}.
Scaling means that the network looks the same at all times in units of the Hubble radius $H^{-1}$: the tension straightens the walls and removes closed ones smaller than $H^{-1}$, while larger structures stay frozen until they enter the Hubble radius, so that there is of order one wall per Hubble volume (energy density $\sim\sigma H$), moving at a sizable fraction of $c$.
The field then follows this induced minimum, which moves towards the vacuum of $V$ ($g=0$) as the relic neutrinos dilute due to the redshift and reaches it at $z_c=13.0$, where $\epsilon=1/4$; afterwards the field oscillates slightly about the vacuum. 

Averaged over the Gaussian distribution of $\delta\phi$, CC SNe fail until $z\simeq13.9$, when $|X|$ falls below $0.45\eV^2$ shortly after $\epsilon$ drops below $0.31$, i.e.\ before the field reaches the vacuum; the fraction $S$ of standard CC-SN feedback then rises to $1/2$ at $z\simeq13.1$ and, as the oscillation about the vacuum decays, to $0.9$ at $z\simeq11.3$, almost independently of the initial displacement $\delta\phi_{\rm rms}$ (Sec.~\ref{sec:early}).

\subsection{First galaxies}
\begin{figure*}[t]
\includegraphics[width=\textwidth]{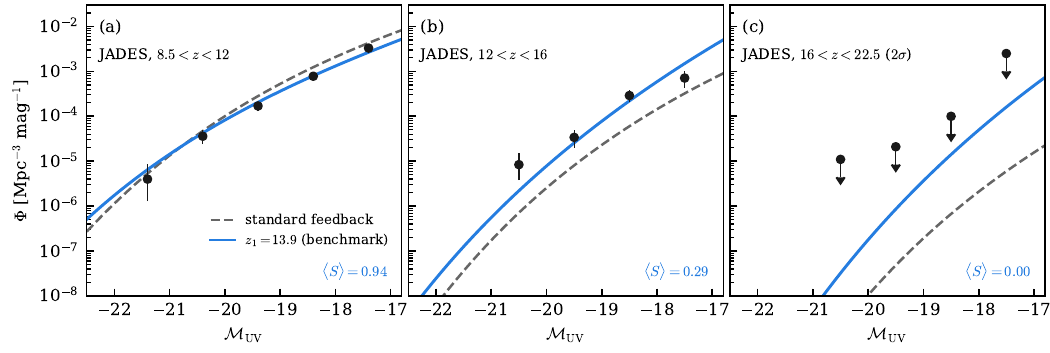}
\caption{\label{fig:uvlf}UV luminosity functions of the three JADES dropout samples~\cite{Whitler2025}, (a) at $8.5<z<12$ and (b) at $12<z<16$, and (c) $2\sigma$ upper limits at $16<z<22.5$, with the best fits for standard feedback ($S=1$, dashed) and for CC SNe resuming at $z_1=13.9$ (solid; $m_1=1.0$~meV for the benchmark). As in the fit, the models are averaged over the redshift range of each sample, weighted by the comoving volume, and $\langle S\rangle$ is the same average of $S$ for the best fit. The fits to PRIMER, to PANORAMIC and to the part of the F150W dropouts at $14<z<16$ are in Fig.~\ref{fig:uvlfscan}(b) and Table~\ref{tab:uvlf}.}
\end{figure*}

\begin{figure*}[t]
\includegraphics[width=\textwidth]{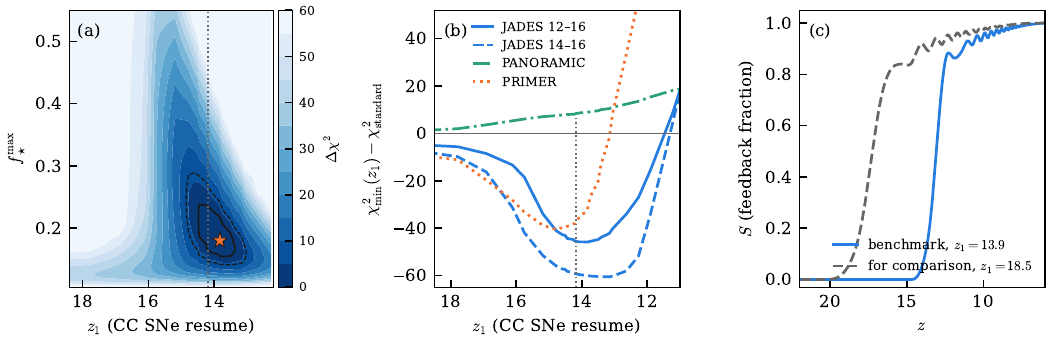}
\caption{\label{fig:uvlfscan}(a) $\Delta\chi^2$ in the plane of the epoch $z_1$ at which CC SNe resume and $f_\star^{\max}$ for JADES at $8.5<z<12$ and $12<z<16$ with its upper limits at $16<z<22.5$; contours: $68\%$ and $95\%$ for two parameters; star: best fit; dotted line: $z_1=14.18$, where oxygen is seen~\cite{Carniani:2024zaf}. (b) $\chi^2$ minimized over $f_\star^{\max}$ as a function of $z_1$, relative to the best standard fit, for each data set of Table~\ref{tab:uvlf} (JADES 12--16 and 14--16: with its points at $12<z<16$ or at $14<z<16$). (c) Fraction $S$ of standard CC-SN feedback, averaged over one period of the field (Sec.~\ref{sec:uvlf}), for the benchmark, $z_1=13.9$, close to the best fit to JADES (solid), and, for comparison, for $z_1=18.5$ (dashed). Here $z_1$ is varied through $m_1$ and, above $14.2$, through $V_{\rm top}$, with the other parameters at the benchmark ($m_1=1.0$~meV, and $m_1=0$ with twice $V_{\rm top}$ in (c)); the other parameters give the same best fits (Fig.~\ref{fig:uvlfparams}).}
\end{figure*}

The relic neutrinos thus keep the first galaxies without CC-SN feedback ($S=0$), everywhere and independently of any wall, until $z_1\simeq13.9$, the epoch at which CC SNe resume, and a heavier lightest neutrino keeps them so for longer, as does a lower $V_{\rm top}$, since $z_1$ is set by $\epsilon$; for comparison, with $m_1=0$ the first explosions would occur at $z_1\simeq14.2$, and with $m_1=0$ and $V_{\rm top}$ doubled already at $z_1\simeq18.5$ [see Fig.~\ref{fig:uvlfscan}(c)]. 
We compute the UV luminosity functions in a semi-empirical model in which haloes form stars at the rate $f_\star f_b\dot M_h$, with an efficiency $f_\star=f_\star^{\max}/(1+S\eta)$ limited by the energy of CC SNe through $\eta=(v_{\rm SN}/v_c)^2$ and calibrated at $z=7$ (Sec.~\ref{sec:uvlf}). The fraction $S$ is obtained from the equation of motion of the scalar field. 
Without feedback $f_\star=f_\star^{\max}$, as in feedback-free starbursts~\cite{Dekel:2023ddd} but in every halo. As found for feedback-free starbursts~\cite{Li:2023xky} and in semi-analytic models with negligible feedback at $z>10$~\cite{Cantarella2025}, suppressing the feedback raises the abundance of bright galaxies; here the suppression and its end follow from the equation of motion of the field, and the luminosity functions of JADES~\cite{Whitler2025} fix the end at $z_1\simeq14$ (Figs.~\ref{fig:uvlf} and~\ref{fig:uvlfscan}; Sec.~\ref{sec:uvlf}).
The other parameters (the tension, the mass of the scalar, the mixing and $\delta\phi_{\rm rms}$) change only the $m_1$ and $V_{\rm top}$ that give $z_1\simeq14$, $m_1=1.0$~meV for the benchmark. The trajectory with $z_1\simeq14$ fits JADES better than standard feedback by $\Delta\chi^2=46$, i.e.\ nominally $6.8\sigma$ for the one additional parameter $z_1$ ($3.8\sigma$ with a model uncertainty of $0.2$~dex; Sec.~\ref{sec:uvlf} and Fig.~\ref{fig:uvlfparams}).
On the other hand, PANORAMIC~\cite{Weibel2025} alone is fitted at least as well with standard feedback ($S=1$ throughout) [Fig.~\ref{fig:uvlfscan}(b)], and with momentum-driven feedback ($\eta=v_{\rm SN}/v_c$) both fits to JADES are poor (Sec.~\ref{sec:uvlf}).
The oxygen seen at $z=14.18$~\cite{Carniani:2024zaf} can come from pair-instability SNe~\cite{Heger:2001cd,Bromm:2003hy,Wise:2010zi}, or from the first CC SNe, since the fit also allows $z_1\simeq14.2$ (Sec.~\ref{sec:uvlf}).
Progenitors of $8$--$10\,\Msun$, about a quarter of CC SNe for a Salpeter mass function~\cite{Salpeter:1955it}, explode easily~\cite{Kitaura:2005bt,Melson:2015tia} and might survive the mixing; if they did, JADES would still be fitted much better than with standard feedback, with a later return of the other explosions, $z_1\simeq12$, and the oxygen could come from the surviving SNe (Sec.~\ref{sec:uvlf}).

\subsection{Walls today, galactic response and synchronicity}
\begin{figure*}[t]
\includegraphics[width=\textwidth]{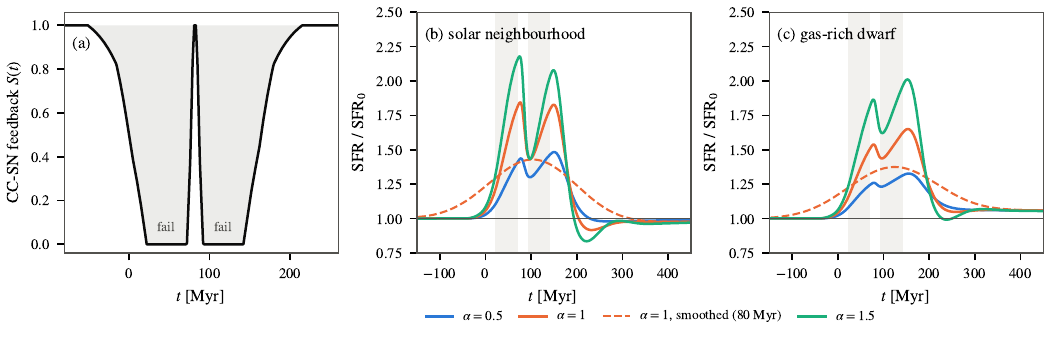}
\caption{\label{fig:response}(a) Fraction $S$ of standard core-collapse-SN feedback delivered while the benchmark wall crosses a galaxy, mapped from the simulations of Ref.~\cite{Mori:2025lky} ($v_w=0.2c$; $t=0$ is where $S$ first drops below $1/2$). (b) Response of the vertically integrated self-regulation model for the solar neighbourhood, for star-formation laws $\mathrm{SFR}\propto\Sigma\rho^{\alpha}$; the dashed curve is the $\alpha=1$ case convolved with a Gaussian age error of $80\Myr$. (c) The same for a gas-rich dwarf. Shading marks the two intervals with $S=0$.}
\end{figure*}

After $z_c$ the mixing is on only inside the walls, which form a network close to scaling.
When a wall passes a galaxy, its CC SNe stop for about $0.1\Gyr$ [Fig.~\ref{fig:response}(a)].
Without their feedback the gas disc loses part of the pressure that supports it and forms stars faster: in a model of the disc in the solar neighbourhood (Sec.~\ref{sec:ism}) the star-formation rate rises by a factor of about two for about $0.1\Gyr$, and a gas-rich dwarf galaxy responds in the same way [Fig.~\ref{fig:response}(b,c)].

A wall that passed the Local Group about $2\Gyr$ ago would thus raise the star formation of all its gas-rich galaxies at the same time, which could explain why their star-formation histories peak at that time; the observed peaks have similar heights but last several times longer.

The wall crosses the whole Local Group within about $10$--$20\Myr$.
The peaks it produces should therefore have the same age in every galaxy, while peaks produced by encounters between galaxies would have different ages.
The present ages of the peaks, obtained with different methods, are not precise enough to tell whether the peaks are simultaneous; this may become possible in the future if all of them are measured in the same way.

For a domain-wall network a given point is crossed within the past $2\Gyr$ with a probability of order $1\%$, which is not much larger than the probability that the peaks in star formation about $2\Gyr$ ago in several galaxies coincide by chance. This comparison will improve with future measurements, and a better understanding of the star-formation histories is essential to test whether the prediction of our scenario is favoured. 

\section{Predictions}

\paragraph*{Surfaces with fewer core-collapse supernovae.}
A galaxy crossed by a wall has fewer CC SNe for about $0.1\Gyr$, while its SNe~Ia, which do not depend on neutrinos, continue, and a gas-rich galaxy forms stars faster, by up to a factor of about two [Fig.~\ref{fig:response}(b,c)].
Since we see distant galaxies as they were in the past, every wall of the network appears on the sky as a thin surface on which CC SNe are reduced and gas-rich galaxies grow faster; whether or not a wall crossed us, such surfaces lie in front of a few per cent of the sky out to $z=0.15$ and of about a tenth out to $z=0.5$.

\begin{figure}[t]
\includegraphics[width=\columnwidth]{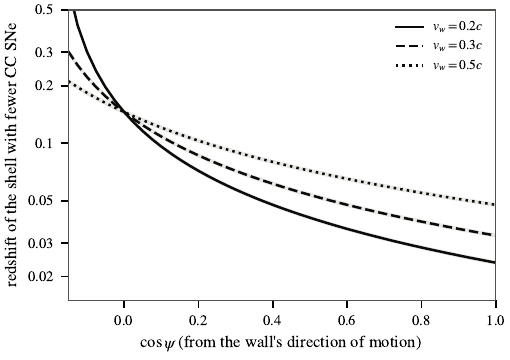}
\caption{\label{fig:shell}Redshift of the shell on our past light cone in which core-collapse supernovae are reduced, as a function of the angle $\psi$ between the line of sight and the wall's motion, for a passage $1.95\Gyr$ ago and $100\Myr$ of failure (flat $\Lambda$CDM), with $v_w=0.2c$ (solid), $0.3c$ (dashed) and $0.5c$ (dotted). The bands show the shell thickness; for the benchmark the $20\Myr$ core splits it in two. No shell appears for $\cos\psi<-v_w/c$.}
\end{figure}

If a wall crossed the Local Group about $2\Gyr$ ago, as assumed above for the peaks of star formation, one of these surfaces is close to us: a thin shell at $z\simeq0.02$ in the direction of motion of the wall and $z\simeq0.15$ perpendicular to it (Fig.~\ref{fig:shell}).
We searched for this shell in three public supernova samples~\cite{Perley:2020ajb,TNS,Barbon:1999dk}, predicting the number of CC SNe in it from the SNe~Ia at the same redshifts (Sec.~\ref{sec:shell}).
The samples show no significant deficit (global $p=0.47$; the largest deficit of the BTS sample alone is not seen in the other two samples) and exclude at $95\%$ confidence $59\%$ of the possible geometries of a wall that passed $1.5$--$2.5\Gyr$ ago, so that the model is not excluded.
LSST~\cite{LSST:2008ijt}, with host-galaxy redshifts, would find $2\times10^3$--$5\times10^3$ CC SNe in the shell, enough for a decisive test. The faster growth of the galaxies in the shell can also be searched for in their star-formation rates from the H$\alpha$ line, as measured in the Sloan Digital Sky Survey~\cite{SDSS:2000hjo,Brinchmann:2003db} and observed in the DESI Bright Galaxy Survey~\cite{Hahn:2022dnf}.

\paragraph*{Early supernovae.}
Neutrino-driven CC SNe, and with them their feedback and metal yields, are absent before $z\simeq13.9$ (before $z\simeq14.2$ even for $m_1=0$) and weakened until $z\simeq11$.

\section{Cosmic birefringence}\label{sec:biref}
The field $\phi$ in our model is consistent with an axion-like particle coupled to photons,
\begin{equation}
\mathcal L\supset c_\gamma\frac{\alpha}{4\pi}\frac{\phi}{f}F_{\mu\nu}\tilde F^{\mu\nu},
\label{eq:photon}
\end{equation}
with a coefficient $c_\gamma$ of order one.
This coupling plays no role in the scenario above and leaves the fine-structure constant $\alpha$ unchanged; its strength, $c_\gamma\alpha/(\pi f)\simeq1\times10^{-17}c_\gamma\,\mathrm{GeV}^{-1}$, is far below laboratory and astrophysical limits.
It rotates the plane of linear polarization of light by $\beta=c_\gamma(\alpha/2\pi)(\phi_{\rm obs}-\phi_{\rm em})/f$, which depends only on the field at emission and at observation~\cite{Carroll:1989vb,Harari:1992ea,Carroll:1998zi,Takahashi:2020tqv}.
When the CMB was emitted the field had just left the top of $V$ (rms distance $1.1f$, symmetric about the top), and today it is in a vacuum, $\pi f/2$ from the top, so that the CMB polarization is rotated by $\pm c_\gamma\alpha/4=\pm0.105^\circ c_\gamma$ on average, with an anisotropic part of rms $0.072^\circ c_\gamma$, comparable to present limits~\cite{Kramer:2026szv}. For $10^{-8}f\lesssim\delta\phi_{\rm rms}\lesssim5\times10^{-4}f$, up to the isocurvature limit (Sec.~\ref{sec:early}), the anisotropic part is $0.03$--$0.07^\circ c_\gamma$; it is below $0.003^\circ c_\gamma$ only for $\delta\phi_{\rm rms}\lesssim10^{-9}f$, where the field is still at the top when the CMB is emitted.
The rotation is the same for any $f$, $m$ and $\sigma$ and in either vacuum, whether or not the wall has crossed us.
The rotation measured in a joint analysis of ACT and Planck, $0.277^\circ\pm0.057^\circ$~\cite{Eskilt:2026imm}, corresponds to $|c_\gamma|=2.65\pm0.55$; $c_\gamma=2$ and $3$ lie within $1.2\sigma$.
The polarization generated at reionization is rotated by $0$ or $\pm c_\gamma\alpha/2$ relative to ours, depending on the vacuum of the scattering region: a large-angle anisotropic signal with an rms of $0.105^\circ c_\gamma$, correlated over about the Hubble radius at reionization~\cite{Kitajima:2022jzz}, which the limit on anisotropic birefringence from the recombination signal~\cite{Kramer:2026szv} does not constrain.
Likewise, light that left a source before the wall crossed it, i.e.\ from beyond the shell, is rotated by $c_\gamma\alpha/2=0.21^\circ c_\gamma$ relative to light from inside it.

\section{Conclusions}
Core-collapse supernovae, which are powered by neutrinos, regulate star formation, and simulations of the progenitor of SN~1987A find that it fails to explode if the electron neutrino mixes with an eV-scale sterile neutrino.
We have pointed out that a light scalar field controlling this mixing can speed up star formation without injecting energy.
In a minimal model the relic neutrinos keep the mixing on until $z\simeq14$, which explains the excess of UV-bright galaxies found by JADES, and a domain wall of the scalar crossing the Local Group would raise the star formation of all its gas-rich galaxies at once.
The model predicts no core-collapse supernovae before $z\simeq14$ and weakened ones until $z\simeq11$, and thin surfaces on our past light cone with fewer core-collapse supernovae and faster star formation, and with a coupling to photons it can explain the hint of isotropic cosmic birefringence.

\begin{acknowledgments}
This work is supported by JSPS KAKENHI Grant Nos.~22K14029 (W.Y.), 23K22486 (W.Y.), 26K00695 (W.Y.) and 26K17199 (D.T.), the Graduate Program on Physics for the Universe (Y.N.), and JST SPRING, Grant Number JPMJSP2114 (Y.N.).
W.Y. is also supported by the Incentive Research Fund of Tokyo Metropolitan University.
W.Y. and D.T. are also supported by the TMU research fund for young scientists.
We acknowledge the academic versions of Claude (Anthropic) and ChatGPT (OpenAI), which provided assistance with literature searches, with coding of the numerical calculations and with writing. The authors conceived the mechanism, designed the study and the model, checked all results and take full responsibility for the content.
\end{acknowledgments}

\clearpage
\onecolumngrid
\setcounter{section}{0}\setcounter{figure}{0}\setcounter{table}{0}\setcounter{equation}{0}
\renewcommand{\thesection}{S\arabic{section}}\renewcommand{\theHsection}{S\arabic{section}}
\renewcommand{\thetable}{S\arabic{table}}\renewcommand{\theHtable}{S\arabic{table}}
\renewcommand{\thefigure}{S\arabic{figure}}\renewcommand{\theHfigure}{S\arabic{figure}}
\renewcommand{\theequation}{S\arabic{equation}}\renewcommand{\theHequation}{S\arabic{equation}}
\begin{center}\textbf{\large Supplemental Material}\end{center}

\section{UV luminosity functions at \texorpdfstring{$z\gtrsim9$}{z > 9}}\label{sec:uvlf}
\paragraph*{Model.}
We fit the UV luminosity function $\Phi$, the comoving number density of galaxies per unit absolute UV magnitude $\mathcal{M}_{\rm UV}$ (AB system).
Each halo hosts one galaxy, whose magnitude $\mathcal{M}_{\rm UV}(M_h,z)$ depends on the halo mass $M_h$ and the redshift $z$, so that
\begin{equation}
\Phi(\mathcal{M}_{\rm UV},z)=\frac{dn}{dM_h}\left|\frac{dM_h}{d\mathcal{M}_{\rm UV}}\right|,
\label{eq:S-lf}
\end{equation}
where $dn/dM_h$ is the comoving number density of haloes per unit halo mass.
For $dn/dM_h$ we use the mass function of Ref.~\cite{Reed:2006rw} for friends-of-friends haloes, with the linear power spectrum of Ref.~\cite{Eisenstein:1997ik}, in the Planck 2018 cosmology~\cite{Planck:2018vyg}: $h=0.6766$, $\Omega_m=0.3111$, $\Omega_b=0.0490$, $n_s=0.9665$ and $\sigma_8=0.8102$.
With COLOSSUS~\cite{Diemer:2017bwl} instead, the values of $\chi^2$ in Table~\ref{tab:uvlf} change by less than $4\%$, and $\Delta\chi^2$ for JADES by $0.3$.
The halo grows at the mean rate of Ref.~\cite{Fakhouri:2010st},
\begin{equation}
\dot M_h=46.1\,\Msun\,{\rm yr}^{-1}\Big(\frac{M_h}{10^{12}\Msun}\Big)^{1.1}(1+1.11z)\sqrt{\Omega_m(1+z)^3+\Omega_\Lambda},
\label{eq:S-mdot}
\end{equation}
with $\Omega_\Lambda\simeq1-\Omega_m$, which we integrate back in time to $z=40$ to obtain the mass history of a halo of mass $M_h$ at $z$.
At each time the halo forms stars at the rate
\begin{equation}
\dot M_\star=f_\star f_b\dot M_h,\qquad f_\star=\frac{f_\star^{\max}}{1+S\eta},\qquad \eta=\Big(\frac{v_{\rm SN}}{v_c}\Big)^2,
\label{eq:S-sfr}
\end{equation}
if its mass exceeds the atomic-cooling mass $M_{\rm cool}=10^8\Msun\,[(1+z)/10]^{-3/2}$, and none otherwise.
Here $f_b=\Omega_b/\Omega_m$ is the cosmic baryon fraction, $f_\star$ the star-formation efficiency, $f_\star^{\max}$ its value without feedback, $S$ the fraction of standard CC-SN feedback at that time, and $v_c=\sqrt{GM_h/R_{\rm vir}}$ the circular velocity at the virial radius $R_{\rm vir}$, defined by $M_h=(4\pi/3)\,18\pi^2\rho_cR_{\rm vir}^3$ with the critical density $\rho_c=3H^2/(8\pi G)$.
The constant velocity $v_{\rm SN}$ sets the strength of the feedback; for each $f_\star^{\max}>0.1$ it is fixed by $f_\star=0.1$ for $M_h=10^{11}\Msun$ at $z=7$ with $S=1$, i.e.\ $\eta=f_\star^{\max}/0.1-1$ there.
For momentum-driven feedback $\eta=v_{\rm SN}/v_c$, with the same calibration.
Standard feedback has $S=1$; in the minimal model $S(z)$ is the history of Sec.~\ref{sec:early}, averaged over one period of the field, centred on each time, and over the eight samples of $\delta\phi$; this average starts the rise of $S$ about $20\Myr$ before $z_1$.
The UV luminosity per unit frequency at cosmic time $t$ is
\begin{equation}
L_{\rm UV}(t)=\int_{t_{40}}^{t}\dot M_\star(t')\,K(t-t')\,dt',\qquad K(\tau)=\begin{cases}K_0\min(1,\tau_0/\tau), & \tau<1\Gyr,\\ 0, & \tau\geq1\Gyr,\end{cases}
\label{eq:S-luv}
\end{equation}
where $t_{40}$ is the cosmic time at $z=40$, $K(\tau)$ the UV luminosity per unit frequency and per unit stellar mass of stars of age $\tau$, and $\tau_0=3\Myr$.
The constant $K_0$ is fixed by $\int_0^{100\,{\rm Myr}}K\,d\tau=1/\kappa_{\rm UV}$ with $\kappa_{\rm UV}=1.15\times10^{-28}\Msun\,\mathrm{yr}^{-1}/(\mathrm{erg\,s^{-1}\,Hz^{-1}})$~\cite{Madau:2014bja}, so that a constant star-formation rate sustained for $100\Myr$ gives $L_{\rm UV}=\dot M_\star/\kappa_{\rm UV}$.
The magnitude is $\mathcal{M}_{\rm UV}=51.63-2.5\log_{10}[L_{\rm UV}/(\mathrm{erg\,s^{-1}\,Hz^{-1}})]$; we include no dust.

\paragraph*{Data and likelihood.}
We fit JADES~\cite{Whitler2025}, PRIMER~\cite{Donnan2024} and PANORAMIC~\cite{Weibel2025} separately, because they share fields, with
\begin{equation}
\chi^2=\sum_{\rm detections}\frac{[\log_{10}(\langle\Phi\rangle/\Phi_{\rm obs})]^2}{\sigma_{\rm d}^2+\sigma_{\rm m}^2}+2\sum_{\rm empty\ bins}N_{\rm exp},
\label{eq:S-chi2}
\end{equation}
where $\Phi_{\rm obs}$ is a measured value and $\langle\Phi\rangle$ is Eq.~(\ref{eq:S-lf}) at the magnitude of that data point, averaged over its redshift range with the comoving volume per unit redshift, $\propto D_c^2/H$ ($D_c$: comoving distance), as weight.
The error $\sigma_{\rm d}$ is the upper error of the data point in dex if the model lies above it and the lower error otherwise, where errors $\Delta_\pm$ given in linear units are converted into $\log_{10}(1+\Delta_+/\Phi_{\rm obs})$ and $-\log_{10}(1-\Delta_-/\Phi_{\rm obs})$, or $1$ if $\Delta_-\geq\Phi_{\rm obs}$; $\sigma_{\rm m}$ is an optional model uncertainty.
In a bin without detections, $N_{\rm exp}=3.8\langle\Phi\rangle/\Phi_{\rm lim}$ is the number of galaxies that the model predicts, since the $2\sigma$ upper limit $\Phi_{\rm lim}$ corresponds to $3.8$ galaxies~\cite{Gehrels:1986mj}, and $2N_{\rm exp}$ is $-2\ln$ of the Poisson probability of no detection.
The redshift ranges are $8.5$--$12$, $12$--$16$ and $16$--$22.5$ for JADES, whose points at $14<z<16$ come from the same galaxies as those at $12<z<16$ and replace them in one of the fits; $8.5$--$9.5$, $9.5$--$10.5$, $10.5$--$11.5$, $11.5$--$13.5$ and $13.5$--$15.5$ for PRIMER; and $8.6$--$11.3$, $11.3$--$15.3$ and $15.3$--$21.9$ for PANORAMIC.
Standard feedback has one free parameter, $f_\star^{\max}$.
The minimal model has $f_\star^{\max}$ and the epoch $z_1$: we vary $m_1$ of the benchmark (Table~\ref{tab:relic}, first row) from $0$ to $14$~meV, which moves $z_1$ from $14.2$ to $8.0$, and, for earlier epochs, $V_{\rm top}$ up to twice its benchmark value at $m_1=0$, which moves $z_1$ up to $18.5$, with the other parameters fixed.
We minimize $\chi^2$ over $f_\star^{\max}$ on a grid from $0.105$ to $0.6$ (steps of $0.005$ below $0.2$ and $0.01$ above), and the $68\%$ and $95\%$ ranges of $z_1$ are where $\chi^2$ exceeds its minimum by less than $1$ and $4$.

\begin{figure}[!htbp]
\centering
\includegraphics[width=0.9\textwidth]{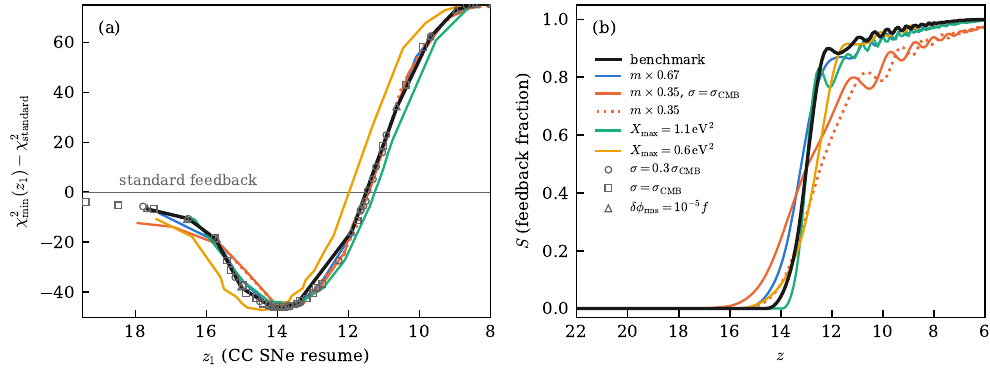}
\caption{\label{fig:uvlfparams}(a) $\chi^2$ of JADES, minimized over $f_\star^{\max}$ and relative to the best standard fit, against the epoch $z_1$ at which CC SNe resume, when one parameter of the benchmark is changed and $m_1$ is varied: the tension (circles, squares), $\delta\phi_{\rm rms}=10^{-5}f$ (triangles), the mass of the scalar ($m\times0.67$ for a failure time of $150\Myr$, $m\times0.35$ for $v_w=0.5c$, solid at $\sigma=\sigma_{\rm CMB}$ and dotted at the benchmark tension) and the mixing ($X_{\max}=1.1$ or $0.6\eV^2$ instead of $0.78\eV^2$, where $X_{\max}$ is $|X|$ at $|g|=1$). Where $m_1=0$ does not reach $z_1\simeq14$, $V_{\rm top}$ is raised at $m_1=0$. (b) $S$ at the best fit of each case.}
\end{figure}

\paragraph*{Results.}
Table~\ref{tab:uvlf} lists the fits.
For JADES the epoch is $z_1=13.8$ ($13.7$--$14.2$ at $68\%$), and the fit depends only on $z_1$: the other parameters change only the $m_1$ and $V_{\rm top}$ that reach it (Fig.~\ref{fig:uvlfparams}).
Standard feedback is recovered at large $z_1$: with $m_1=0$ and $V_{\rm top}$ raised by factors of $4$--$16$, $z_1=24$--$41$ and $\chi^2=55.8$--$56.6$.
We draw $20\,000$ mock JADES data sets from the best standard fit, with the errors of the data and the bins without detections as observed, and fit both models to each, including $z_1=24$--$41$; an improvement above $\Delta\chi^2=9$ occurs in $0.16\%$ of them, about half as often as for $\chi^2$ with one degree of freedom.
The improvement for JADES remains, $\Delta\chi^2=43$, when the calibration value $0.1$ is varied from $0.07$ to $0.14$ and a Gaussian scatter of $0.5$~mag in $\mathcal{M}_{\rm UV}$ at fixed $M_h$ is allowed, in both models; PRIMER is fitted poorly by both models.
Without the average over one period of the field the best fit has $\chi^2=12.0$ and the $68\%$ range of $z_1$ is $13.7$--$14.5$; with an average over the preceding period instead, $\chi^2=11.6$ and $14.2$--$14.8$, and the benchmark, $z_1\simeq13.9$, is inside the $95\%$ range.
If a quarter of CC SNe still explode, i.e.\ $S$ is replaced by $1/4+3S/4$, the best fit to JADES, at $z_1\simeq11.9$, has $\chi^2=13.1$, against $57.5$ for standard feedback.

\begin{table}[h]
\caption{\label{tab:uvlf}Best fits to the UV luminosity functions, each survey separately ($13$, $11$, $10$ and $29$ data points): $\chi^2$, and in brackets with a model uncertainty of $0.2$~dex. For the minimal model $z_1$ is fitted ($13.8$, $13.2$, $18.5$ and $14.8$ for energy-driven feedback; $18.5$, for $m_1=0$ and twice $V_{\rm top}$, is the earliest in the scan); the row $z_1=14.2$ is for $m_1=0$. Calibration $f_\star=0.1$, no scatter.}
\begin{ruledtabular}
\begin{tabular}{lcccc}
 & JADES ($12<z<16$) & JADES ($14<z<16$) & PANORAMIC & PRIMER \\ \hline
Standard feedback & $57.5$ ($20.8$) & $71.7$ ($45.1$) & $6.4$ ($3.7$) & $95.3$ ($46.7$) \\
Minimal model & $11.5$ ($6.4$) & $11.1$ ($8.5$) & $7.8$ ($4.7$) & $54.7$ ($25.0$) \\
\quad $z_1=14.2$ & $12.1$ ($7.2$) & $12.7$ ($9.2$) & $14.5$ ($9.0$) & $57.1$ ($25.0$) \\
Momentum-driven, standard & $58.2$ ($21.3$) & $73.4$ ($46.2$) & $8.3$ ($4.6$) & $137.2$ ($53.2$) \\
Momentum-driven, minimal model & $29.1$ ($11.4$) & $27.4$ ($11.9$) & $13.4$ ($7.8$) & $139.5$ ($48.3$)
\end{tabular}
\end{ruledtabular}
\end{table}

\section{Self-regulation model and the star-formation histories}
\label{sec:ism}
\begin{figure}[!htbp]
\centering
\includegraphics[width=0.6\textwidth]{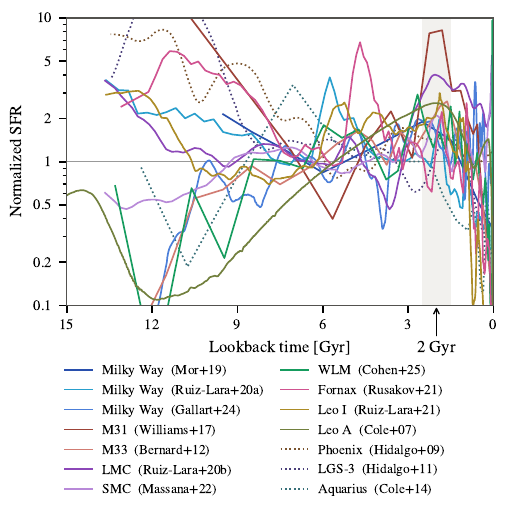}
\caption{\label{fig:sfh}Reconstructed star-formation rates (SFRs) as a function of look-back time for the Milky Way~\cite{Mor2019,RuizLara2020,Gallart2024}, M31~\cite{Williams2017}, M33~\cite{Bernard:2011vg}, the LMC~\cite{RuizLara2020b}, the SMC~\cite{Massana2022}, WLM~\cite{Cohen2025}, Fornax~\cite{Rusakov2021}, Leo~I~\cite{RuizLara2021LeoI}, Leo~A~\cite{Cole:2007ac}, Phoenix~\cite{Hidalgo:2009kf}, LGS-3~\cite{Hidalgo:2011wb}, and Aquarius~\cite{Cole2014}. Each reconstruction is normalized to the time-weighted 30th percentile of its SFR over the past $7.5\Gyr$. The first nine galaxies (solid lines) show a peak around $2\Gyr$ ago (marked by an arrow), whereas the remaining three galaxies (dotted lines) do not. The selection is illustrative: about half of the Local Group dwarfs that formed stars recently show such a peak.}
\end{figure}

\paragraph*{Star-formation histories in the Local Group.}
Peaks of the SFR about $2\Gyr$ ago have been reported in the solar neighbourhood~\cite{RuizLara2020,Mor2019}, in the Magellanic Clouds~\cite{RuizLara2020b,Massana2022}, in M31 and M33~\cite{Bernard:2011vg,Williams2017} and in roughly half of the Local Group dwarfs that formed stars recently (Fig.~\ref{fig:sfh}), and surveys of dwarf galaxies find rising or peaked histories over the past $2$--$3\Gyr$ in a non-negligible fraction of them~\cite{Olsen2021,Weisz:2011gp,Weisz:2014qra}.
They are usually attributed to processes specific to each galaxy or pair, such as passages of the Sagittarius dwarf~\cite{RuizLara2020}, interactions within the LMC--SMC and M31--M33 pairs~\cite{Massana2022,Bernard:2011vg}, a major merger for M31~\cite{Hammer2018}, mergers or interactions for Fornax~\cite{Rusakov2021} and the infall of Leo~I into the Milky Way halo~\cite{Bennet2024}.
Both the apparent coincidence of the peaks and their absence in some galaxies could arise from systematic age offsets due to the choice of fitting method and stellar-evolution model~\cite{Dolphin:2012er,Williams2017}.
A perturbation that affected galaxies throughout the Local Group could explain contemporaneous peaks with a single event; a passing wall is such an event, with a probability of order $1\%$ within the past $2\Gyr$.

M31 formed $\sim60\%$ of the stars of the past $5\Gyr$ between about $2$ and $4\Gyr$ ago~\cite{Williams2015}, far more than a passing wall can produce, so that this episode needs its own cause, for example a merger $\sim2$--$3\Gyr$ ago~\cite{DSouzaBell2018,Hammer2018}.
A wall could still have set when it began: if the gas brought in by the merger was close to collapse, $0.1\Gyr$ without supernova feedback could start a collapse that continues after the supernovae return.
If the wall did trigger this episode, the episode must have begun after the wall passed.

\paragraph*{Galactic response.}
We model a disc annulus as a gas layer of surface density $\Sigma$ in the potential of stars and dark matter, whose weight is carried by the thermal pressure, tied to the far-ultraviolet-weighted SFR~\cite{Ostriker:2010zc}, by turbulence driven by the momentum of CC SNe (delayed by $3$--$40\Myr$ and multiplied by $S$), of SNe~Ia and of pre-SN feedback~\cite{Ostriker:2011jb}, by a magnetic field and by cosmic rays.
The layer relaxes to equilibrium in a crossing time and forms stars at $\mathrm{SFR}\propto\Sigma\rho_{\rm mid}^{\alpha}$, where $\rho_{\rm mid}$ is the midplane density, with $\alpha=0.5$ (free fall with constant efficiency~\cite{Kim:2013xaa}) to $1.5$; calibrated to the solar neighbourhood, it requires feedback yields consistent with simulations~\cite{KimOstriker2015,OstrikerKim2022}.
When the benchmark wall passes, the SFR rises in two peaks $\simeq75\Myr$ apart, separated by the core of the wall, and the delayed SNe of the new massive stars cause a dip afterwards [Fig.~\ref{fig:response}(b) of the main text]; a gas-rich dwarf responds in the same way [Fig.~\ref{fig:response}(c)].
Table~\ref{tab:peak} compares the peak with the reconstructions of Fig.~\ref{fig:sfh}: the heights are similar, while the observed peaks last longer, about twice as long for the solar neighbourhood of Ruiz-Lara et al.~\cite{RuizLara2020} and several times to an order of magnitude longer in the other reconstructions, which age errors may have broadened.
The response is limited because the layer thins when it loses support, which reduces its weight; at fixed weight~\cite{OstrikerKim2022} the SFR would rise by $\simeq3.7$, and hydrodynamical simulations without CC-SN feedback give enhancements from negligible to about an order of magnitude~\cite{Forbes2016,Hopkins:2017ycn,Shimizu2019,Whitworth2026}, depending on how the feedback couples to dense gas~\cite{Hennebelle:2014ska}.
Nevertheless, SN feedback can also trigger star formation by compressing gas through expanding shells~\cite{Inutsuka2015,Zucker2022,Herrington2023}; thus, whether its net effect on the SFR is negative remains open to debate.

\begin{table}[t]
\caption{\label{tab:peak}The SFR peak about $2\Gyr$ ago in the model and in the reconstructions of Fig.~\ref{fig:sfh}. Height: maximum SFR between $1.5$ and $2.5\Gyr$ ago divided by the baseline; duration: area above the baseline in the same window divided by the baseline. The baseline is the SFR before the wall in the model and the 30th percentile of the SFR over the last $7.5\Gyr$ in the reconstructions.}
\begin{ruledtabular}
\begin{tabular}{p{0.47\columnwidth}p{0.26\columnwidth}p{0.17\columnwidth}}
 & Height & Duration [Gyr] \\ \hline
Model, solar neighbourhood ($\alpha=0.5$, $1$, $1.5$) & $1.48$, $1.85$, $2.18$ & $0.06$--$0.13$ \\
\quad a quarter of CC SNe explode, or failure only at $|X|=0.78\eV^2$ & $1.31$--$1.98$ & \\
\quad all of $8$--$12\,\Msun$ explode ($43\%$) & $1.20$--$1.51$ & \\
Model, gas-rich dwarf & $1.33$--$2.01$ & $0.06$--$0.14$ \\
Milky Way, Ruiz-Lara et al.~\cite{RuizLara2020} & $\sim2$ & $\sim0.2$ \\
Milky Way, Refs.~\cite{Mor2019,Gallart2024} & $\sim2$ & $0.5$--$0.7$ \\
Other eight galaxies with a peak (median) & $\sim3$ & $1.1$
\end{tabular}
\end{ruledtabular}
\end{table}

\paragraph*{Model details.}
State variables: gas surface density $\Sigma$, layer half thickness $h$, turbulent momentum $\Sigma\sigma_t$, cosmic-ray pressure $P_{\rm CR}$ and turbulent magnetic pressure $P_{B,t}$.
With $\rho=\Sigma/(2h)$:
\begin{align}
\Sigma_{\rm SFR}&=\Sigma_{\rm SFR,0}\,\frac{\Sigma}{\Sigma_0}\Big(\frac{\rho}{\rho_0}\Big)^{\alpha},\\
\frac{d(\Sigma\sigma_t)}{dt}&=\frac12\Big[p_{\rm CC}\,S(t)\,R_{\rm CC}(t)+p_{\rm Ia}R_{\rm Ia}+p_{\rm early}\Sigma_{\rm SFR}\Big]-\frac{\Sigma\sigma_t^2}{h},\\
P_{\rm th}&=\Upsilon_{\rm th}\,\langle\Sigma_{\rm SFR}\rangle_{\rm FUV}\quad(\text{disc}),\qquad P_{\rm th}=\rho c_w^2\quad(\text{dwarf}),\\
\frac{dh}{dt}&=\frac{h_{\rm eq}-h}{h/\sigma_{\rm eff}},\qquad \tfrac\pi2G\Sigma^2+\pi G\rho_{\rm sd}\Sigma h_{\rm eq}=P_{\rm tot}(h).
\end{align}
All pressures are evaluated at the current thickness $h$, so that $h_{\rm eq}$ is the thickness at which the weight would balance them, $P_{\rm tot}=P_{\rm th}+P_{\rm turb}+P_{\rm mag}+P_{\rm CR}$ with $P_{\rm turb}=\rho\sigma_t^2$, and $\sigma_{\rm eff}=(P_{\rm tot}/\rho)^{1/2}$; $\rho_{\rm sd}$ is the midplane density of stars and dark matter.
Here $R_{\rm CC}$ is the SFR convolved with a uniform delay distribution over $3$--$40\Myr$, and $\langle\cdot\rangle_{\rm FUV}$ a convolution with an FUV light curve (flat to $3\Myr$, then $\propto t^{-1.25}$); $P_{\rm CR}$ relaxes to the SN rate on $20\Myr$ and $P_{B,t}$ to $(1-f_{\rm mean})P_{\rm mag,0}P_{\rm turb}/P_{\rm turb,0}$ on $10\Myr$, where the subscript $0$ marks the baseline and $P_{\rm mag}=f_{\rm mean}P_{\rm mag,0}+P_{B,t}$ includes a mean field, a fraction $f_{\rm mean}=0.3$, that SN turbulence does not maintain.
The baseline pressure split is $(P_{\rm th},P_{\rm turb},P_{\rm mag},P_{\rm CR})=(0.15,0.40,0.30,0.15)\,W_0$ for the disc; the dwarf has an isothermal warm phase ($c_w=6\,\mathrm{km\,s^{-1}}$), magnetic and CR fractions $0.10$ and $0.05$, and an SN-driven outflow loading $5$.
The momenta $p_{\rm CC}$ and $p_{\rm early}$ are per unit stellar mass formed, $p_{\rm early}=300\,\mathrm{km\,s^{-1}}$, and the SN Ia term $p_{\rm Ia}R_{\rm Ia}$ is constant, $0.25$ (disc) or $0.10$ (dwarf) of the CC term at baseline.
The calibration to the solar neighbourhood uses $\Sigma=13.7\,\Msun\,\mathrm{pc}^{-2}$, $\rho_{\rm sd}=0.056\,\Msun\,\mathrm{pc}^{-3}$~\cite{McKee:2015hwa}, $\Sigma_{\rm SFR}=2.5\times10^{-3}\,\Msun\,\mathrm{yr^{-1}\,kpc^{-2}}$ and a baseline weight $W_0/k_B=2.4\times10^4\,\mathrm{K\,cm^{-3}}$; it requires $p_{\rm CC}\simeq2300\,\mathrm{km\,s^{-1}}$ from CC SNe and a thermal yield $\Upsilon_{\rm th}\simeq300\,\mathrm{km\,s^{-1}}$, consistent with simulations~\cite{KimOstriker2015,OstrikerKim2022}.

\section{Early-universe history with sterile neutrinos}
\label{sec:map}\label{sec:early}
We integrate the equation of motion of the field in flat $\Lambda$CDM, including the sterile states: 
\begin{equation}
\ddot u+3H\dot u+\frac{m^2}{2}\big[\sin2u+2\epsilon_h\sin4u-\epsilon\,s'(u)\big]=0 ,
\label{eq:S-eom}
\end{equation}
where $u=\phi/f$, $\epsilon$ is that of Eq.~(\ref{eq:Veff}) of the main text, $\epsilon_h=\Delta E_h/V_{\rm top}$ accounts for sterile states, and $s(u)\simeq\sin^22u$.
While the field is frozen near the top, non-resonant production~\cite{Dodelson:1993je,Hannestad:2012ky} makes locally a fraction $f_s\simeq484\,(\delta\phi/f)^2$ of a thermal population of sterile states, which raise the energy by $\Delta E_h\simeq f_sn_\nu m_D^2/(2M)=230\,f_s\,\Delta E_\nu$ at $|g|=1$ once nonrelativistic, and by much more while relativistic; since this energy grows with $|g|$, it can make the top a minimum and hold the field there.
This is negligible for the benchmark, $\delta\phi_{\rm rms}=10^{-7}f$ ($f_s\sim5\times10^{-12}$), and up to $\delta\phi_{\rm rms}\simeq2\times10^{-4}f$; for larger $\delta\phi_{\rm rms}$ the sterile states hold the field at the top until after recombination in part of the sky, a quarter of it at $3\times10^{-4}f$ and three eighths at $5\times10^{-4}f$.
The density perturbation of the sterile states, $\propto f_s$, is an isocurvature perturbation of the neutrinos. The walls separate regions of opposite sign of $\delta\phi$, Gaussian with zero mean, about as large as the Hubble radius at $z\sim10^3$, which the CMB resolves; $f_s$ therefore varies on this scale by of order itself, whatever the origin of $\delta\phi$, and comparing this with the Planck limit~\cite{Planck:2018jri} on isocurvature perturbations with a power-law spectrum gives roughly $f_s\lesssim10^{-4}$, i.e.\ $\delta\phi_{\rm rms}\lesssim5\times10^{-4}f$. The other inhomogeneities are negligible. At recombination the energy density of the field, $\simeq V_{\rm top}$, is $2\times10^{-11}$ of the total, and the energy that the field exchanges with the relic neutrinos in the mixed phase, $\lesssim\epsilon V_{\rm top}$ with $\epsilon\simeq7\times10^3$, is at most $\sim10^{-7}$ of it, far below the perturbations of $\sim10^{-5}$ seen in the CMB.
We start at rest at $z=3\times10^6$ with eight displacements, the mid-points of eight equal-probability bins of the Gaussian distribution of $\delta\phi$, each with its own $f_s$, and average the results.
Between $|X|=0.45$ and $0.05\eV^2$, $S$ interpolates linearly through the diagnostic explosion energies of the simulations of the progenitor of SN~1987A, evaluated when the shock reaches $2000$~km, $43\%$, $82\%$ and $93\%$ of that without mixing at $|X|=0.26$, $0.15$ and $0.09\eV^2$~\cite{Mori:2025lky}.
Table~\ref{tab:relic} checks how the history of $S$ depends on the parameters: $z_1$ is the redshift of the first explosions, $z_{1/2}$, $z_{0.9}$ and $z_{0.98}$ are those at which $S$, averaged over one period of the field, reaches $1/2$, $0.9$ and $0.98$, and $z_c$ is that at which the minimum induced by the relic neutrinos reaches the vacuum, $\epsilon=1/4$; the field first reaches it about $20\Myr$ later.
For the benchmark these epochs change by less than $0.3$ for $\delta\phi_{\rm rms}=10^{-8}$--$3\times10^{-4}f$, and by less than $0.4$ up to $5\times10^{-4}f$.

\begin{table}[h]
\caption{\label{tab:relic}History of the field for the benchmark wall ($v_w=0.2c$, failure time $T_{\rm fail}=100\Myr$, $\delta\phi_{\rm rms}=10^{-7}f$), averaged over the eight samples of $\delta\phi$; the first row is the benchmark, and each of the others changes one parameter, for comparison. NO: normal ordering, with Majorana masses.}
\begin{ruledtabular}
\begin{tabular}{llccccccc}
$\sigma/\sigma_{\rm CMB}$ & masses & field leaves the top & $z_c$ & $z_1$ & $z_{1/2}$ & $z_{0.9}$ & $z_{0.98}$ & $|X|$ today [eV$^2$] \\ \hline
0.5 & NO, $m_1=1.0$~meV & $z=1200$--$1530$ & 13.0 & 13.9 & 13.1 & 11.3 & 8.1 & $2.8\times10^{-3}$ \\
1 & NO, $m_1=1.0$~meV & $740$--$960$ & 17.0 & 18.2 & 17.1 & 13.8 & 9.7 & $2.2\times10^{-3}$ \\
0.3 & NO, $m_1=1.0$~meV & $1680$--$2120$ & 10.6 & 11.5 & 10.7 & 9.6 & 7.1 & $3.2\times10^{-3}$ \\
0.5 & NO, $m_1=0$ & $1190$--$1510$ & 13.2 & 14.2 & 13.3 & 11.6 & 8.4 & $2.6\times10^{-3}$
\end{tabular}
\end{ruledtabular}
\end{table}

\section{Shell with fewer core-collapse supernovae}
\label{sec:shell}
\paragraph*{Geometry and forecast.}
A galaxy is inside the shell if the light we receive left it while the wall was stopping its CC SNe.
For a passage $\tau_p$ ago the shell lies at comoving distance $\simeq v_w\tau_p/(v_w/c+\cos\psi)$, where $\psi$ is the angle between the line of sight and the motion of the wall, only for $\cos\psi>-v_w/c$; for $v_w=0.2c$ and $\tau_p\simeq1.95\Gyr$ it lies at $z\simeq0.023$ in the forward direction and $z\simeq0.146$ perpendicular to the motion, with thickness $\Delta z\sim10^{-3}$--$10^{-2}$.
For a scaling network a line of sight crosses on average $\frac12\int\mathcal A\,d\chi$ walls, where $\mathcal A\sim aH$ is the comoving wall area per comoving volume, so that such surfaces lie in front of a few per cent of the sky out to $z=0.15$ and of about a tenth out to $z=0.5$.
With rate densities of $10^{-4}$ (CC) and $2.6\times10^{-5}$ (Ia) $\mathrm{yr^{-1}\,Mpc^{-3}}$ (cf.\ Ref.~\cite{Li:2010kd}), the shell of the benchmark contains about $50$ CC SNe in a survey like the ZTF Bright Transient Survey (BTS)~\cite{Perley:2020ajb} and $2\times10^3$--$5\times10^3$ in LSST with host redshifts~\cite{LSST:2008ijt}.

\paragraph*{Samples and statistic.}
We use the public BTS sample with redshifts quoted to four or more decimals ($1299$ CC and $2433$ Ia at $0.004<z<0.15$, retrieved on 2026 November 29), the classified SNe of the Transient Name Server~\cite{TNS} that are not in the BTS list ($1001$ CC and $1343$ Ia) and the Asiago catalogue~\cite{Barbon:1999dk} ($1632$ CC and $1589$ Ia).
For a front direction, a speed $v_w$, a passage time $\tau_p$ and a duration $T_{\rm fail}$, each SN has a probability $P$ of lying in the shell, and with the CC-to-Ia ratio $R_b$ in redshift bins the deficit of CC SNe is measured by
\begin{equation}
T=\sum_{\rm Ia}R_bP-\sum_{\rm CC}P,\qquad
\mathrm{Var}\,T=\sum_bR_b\,\frac{N_b}{N_b-1}\sum_{i\in b}(P_i-\bar P_b)^2,\qquad Z=\frac{T}{\sqrt{\mathrm{Var}\,T}},
\end{equation}
with expectation $Z_1=\sum_{\rm Ia}R_bP^2/\sqrt{\mathrm{Var}\,T}$ for a real shell; the statistic and the scan range were fixed on mock catalogues before the data were scanned.

\paragraph*{Result.}
The BTS sample alone has its largest deficit, $Z=5.2$, for a wall that passed $1.70\Gyr$ ago at $0.105c$, with $p=0.04$ from $100$ permutations within the sample; at this geometry the Asiago and TNS samples have more CC SNe than predicted, $Z=-1.7$ and $-0.9$, against $1.4$ and $1.3$ expected for a real shell.
The three samples combined have their largest deficit at $Z=4.7$, with $p=0.47$ from $100$ permutations of the CC and Ia labels.

\paragraph*{Exclusion.}
A geometry is excluded at $95\%$ confidence if $\mathrm{CL}_s=F(Z-0.9\,Z_1)/F(Z)<0.05$, where $F$ is the cumulative distribution function of the standard normal distribution and the factor $0.9$ allows for a shell that lies between the points of the scan; shells injected into mock catalogues, and evaluated slightly off their geometry, are excluded in $0$ of $300$ cases.
Table~\ref{tab:excl} gives the excluded fractions, without the weakened explosions on either side of the failure intervals, so that they are conservative; if a fraction $S_{\min}$ of the CC SNe still explodes, $Z_1$ is multiplied by $1-S_{\min}$.
The values $S_{\min}=0.25$ and $0.43$ correspond to the survival of the progenitors of $8$--$10$ and $8$--$12\,\Msun$, which give about a quarter and $43\%$ of CC SNe for a Salpeter mass function~\cite{Salpeter:1955it}.

\begin{table}[h]
\caption{\label{tab:excl}Fraction of wall geometries excluded at $95\%$ confidence by the BTS, TNS and Asiago samples together (all directions, $\tau_p=1.5$--$2.5\Gyr$; last column: $v_w=0.2$--$0.3c$ and $\tau_p=1.7$--$2.2\Gyr$), for a single failure interval and, in the last rows, for the two intervals of the wall of the main text, also with a fraction $S_{\min}$ of the CC SNe still exploding (BTS sample retrieved on 2026 November 29).}
\begin{ruledtabular}
\begin{tabular}{ccccccc}
$T_{\rm fail}$ [Myr] & all $v_w$ & $v_w/c=0.10$--$0.15$ & $0.15$--$0.3$ & $0.3$--$0.6$ & $0.6$--$1$ & benchmark region \\ \hline
100 & 0.63 & 0.77 & 0.75 & 0.61 & 0.40 & 0.77 \\
150 & 0.84 & 0.96 & 0.93 & 0.83 & 0.64 & 0.94 \\
\multicolumn{7}{c}{two failure intervals (main text); $T_{\rm fail}$ is the total failure time} \\
100 & 0.59 & 0.74 & 0.72 & 0.55 & 0.34 & 0.73 \\
150 & 0.80 & 0.94 & 0.90 & 0.79 & 0.59 & 0.91 \\
\multicolumn{7}{c}{two failure intervals with a fraction $S_{\min}$ of the CC SNe still exploding} \\
100, $S_{\min}=0.25$ & 0.34 & 0.42 & 0.46 & 0.31 & 0.16 & 0.48 \\
100, $S_{\min}=0.43$ & 0.14 & 0.15 & 0.22 & 0.13 & 0.05 & 0.24
\end{tabular}
\end{ruledtabular}
\end{table}

\bibliography{references}

\end{document}